\documentclass[onecolumn,secnumarabic,amssymb, amsmath, nofootinbib,tightenlines,
nobibnotes, aps, prl,epsfig]{revtex4}
\usepackage{graphicx}
\usepackage{dcolumn}
\usepackage{bm}

\usepackage{amssymb}
\usepackage{epsfig}
\usepackage{color}

\newcommand{\ba}{\begin{eqnarray}}
\newcommand{\ea}{\end{eqnarray}}
\newcommand{\be}{\begin{equation}}
\newcommand{\ee}{\end{equation}}
\newcommand{\bdisplay}{\begin{displaymath}}
\newcommand{\edisplay}{\end{displaymath}}

\begin{document}
\preprint{APS/123-QED}
\title{A limit on top quark pair production at future electron-proton  colliders}

\author{G. R. Boroun}%
 \email{ boroun@razi.ac.ir }
\affiliation{Department of physics, Razi University, Kermanshah
67149, Iran}

\date{\today}

 \pacs{***}
\begin{abstract}
The ratio of the structure functions for deep inelastic scattering
in top pair production at future electron-proton colliders is
analyzed at a fixed $\sqrt{s}$ and $Q^2$ relative to the minimum
value of $x$ given by $Q^2/s$ using collinear factorization. This
compact formula for the ratio $F_{L2}(Q^2/s,Q^2,m_{t})$ is useful
for extracting a bound on the top structure function. The reduced
cross-section for top production at this limit is determined,
establishing a bound value for $t\overline{t}$ production  at the
LHeC and FCC-eh center-of-mass energies based on renormalization
scales. The modification of the Bjorken scaling is applied to this
bound of the reduced top cross-section at the renormalization
scale $\mu^2=Q^2+4m_{t}^{2}$, which improves the scaling quantity
at $Q^2{<}4m_{t}^{2}$. The dipole cross-section for top pair
production is examined across a wide range of dipole sizes,
denoted as $r$. It is expected that there will be limited behavior
in observing top saturation in future electron-proton colliders
according to the bound behavior. The probability of the Higgs
boson in $\gamma^{*}g$ interactions is compared to the $gg$
process at
the order of $\mathcal{O}(\alpha^{\mathrm{em}}/\alpha_{s})$.\\

\end{abstract}
 \pacs{***}
\maketitle
\section{I. Introduction}

One of the main physics goals of the Large Hadron Electron
Collider (LHeC) and the Future Circular Collider Electron-Hadron
(FCC-eh) is to search for the top quark \cite{LHeC}. These new
colliders, namely LHeC and FCC-eh, increase the center-of-mass
energy. At the LHeC, it can reach approximately
$\sqrt{s}{\simeq}1.3~\mathrm{TeV}$ and at the FCC-eh, it would
reach $\sqrt{s}{=}3.5~\mathrm{TeV}$ with a similar luminosity as
the LHeC. The kinematic range in the LHeC for electron and
positron neutral-current (NC) interactions in the perturbative
region is well below $x{\approx}10^{-6}$ and extends up to
$Q{\simeq}1~\mathrm{TeV}$. This range will be extended down to
$x{\simeq}10^{-7}$ at the FCC-eh option of a Future Circular
Collider program \cite{FCC, Klein}. The top cross section can be
measured through the boson-gluon fusion (BGF) process which allows
for a deeper understanding of the hadronic structure at high
energies through deep inelastic scattering (DIS). One of the
important production modes for top quarks at the LHeC is
photoproduction of top-antitop quark pairs ($t\overline{t}$), for
which a total cross-section of $0.05~\mathrm{pb}$ is expected at
the LHeC. The LHeC and FCC-he colliders will create new research
opportunities with top cross sections that could not be observed
at HERA. The total cross section in DIS for top quark pair
production is approximately $53~\mathrm{fb}$ at the LHeC and about
$1~\mathrm{pb}$ at the FCC-he \cite{BroS}.\\
The top quark is produced in nuclear collisions by gluon-gluon
fusion\footnote{Neutral current (NC) top quark production occurs in both the DIS and photoproduction modes, where single
top and top pair production take place, respectively. In
the LHeC, a combination of a $60~\mathrm{GeV}$ electron
beam and a $7~\mathrm{TeV}$ proton beam from the Large Hadron Collider
(LHC), the DIS top pair cross section is approximately $23~\mathrm{fb}$,
which  is lower than at the LHC, but offers better measurement potential \cite{BroS}. In the LHC, any heavy $s$-channel resonance with a significant branching ratio
to $t\overline{t}$ is introduced by the processes
$gg{\rightarrow}t\overline{t}$,  where $\%15$ of the contributions
to $t\overline{t}$ production come from these processes. One of the
production modes for top quarks at the LHeC is the process
$\gamma^{*}g{\rightarrow}t\overline{t}$, which will explore the
strange density and  momentum fraction carried by top quarks,
something that was previously impossible at HERA. } (GGF) \cite{Gao, Okor, Bylund,
Nocera}, $g+g{\rightarrow}t+\overline{t}$, at the LHC. The top
quark pair production cross sections are
$\sigma_{t\overline{t}}=984.5~ \mathrm{pb}$ at
$\sqrt{s}=14~\mathrm{TeV}$ and $\sigma_{t\overline{t}}=826.4~
\mathrm{pb}$ at $\sqrt{s}=13~\mathrm{TeV}$ at CMS \cite{CMS} and
ATLAS \cite{ATLAS}. Top quark pair production at the LHC is
characterized by final states comprising the decay products of the
two $W$ bosons and two $b$ jets, where the
$\sigma_{tot}^{t\overline{t}}$ at ultra-high energies for
$\mu=m_{t}$ can be written as \cite{Okor}
\begin{eqnarray} \label{Sigmagg eq}
\widehat{\sigma}_{gg}(\beta,m_{t})=\frac{\alpha_{s}^{2}}{m_{t}^{2}}
\{ f_{gg}^{(0)}+\alpha_{s}f_{gg}^{(1)}+\alpha_{s}^{2}f_{gg}^{(2)}+
\mathcal{O}( \alpha_{s}^{3}) \}.
\end{eqnarray}
The functions $f_{gg}$ are known at leading and high-order
approximations and depend only on dimensionless parameters $\beta$
and $\rho$, where $\beta^{2}=1-\rho$ with $\rho=4m_{t}^{2}/s$ as
the squared relative velocity of the final state top quarks having
pole mass $m_{t}$ and produced at the square of the gluonic center
of mass energy $s$.\\
The study of production mechanisms of top pair quarks provides us
with new tests of quantum chromodynamics (QCD) such as Higgs
bosons in $t\overline{t}$ production. The top quark has a large
Yukawa coupling with the Higgs boson. In the usual extensions of
the standard model the Higgs sector includes extra scalars, which
also tend to couple strongly with the top quark. This production
which is a crucial process at the LHC is studied in
Refs.\cite{Higgs1, Higgs2, Higgs3, Higgs4}. This means that the
top quark with a mass of about $172.5{\pm}0.5~\mathrm{GeV}$,
measured by ATLAS \cite{ATLAS} and CMS \cite{CMS}, and a particle
that most strongly influences the Higgs boson and its potential is
special among all quarks and can be considered at future
colliders. The observed effects of the $t\overline{t}$ productions
at the LHC are implemented on the CT18 parton distribution
functions (PDFs)
recently in \cite{CT18}.\\
Several methods have been proposed for determining the top
structure functions in the DIS  that are relevant for
investigating ultra-high energy processes, such as the scattering
of cosmic neutrinos from hadrons  \cite{Boroun1, Boroun2, Boroun3,
Schuler, Baur}. The interaction Lagrangian density due to the
electromagnetic coupling in $ep{\rightarrow}t\overline{t}X$ pair
production \cite{Schuler, Baur} is defined by the following
equation:
\begin{eqnarray} \label{Lagran eq}
\mathcal{L}_{em}=-\sqrt{4\pi
\alpha_{em}}\bigg{[}e_{e}\overline{\Psi}_{e}{\gamma
^{\mu}}\Psi_{e}+e_{t}\overline{\Psi}_{t}{\gamma
^{\mu}}\Psi_{t}\bigg{]}A_{\mu}.
\end{eqnarray}
Here, the Dirac fields $\Psi_{e}$, $\Psi_{t}$, and the vector
field $A_{\mu}$  describe the electron, the top and the photon,
respectively. $e_{e}$ and $e_{t}$ represent the electric charge of
the electron and top-quark in units of $\sqrt{4\pi \alpha_{em}}$,
where $\alpha_{em}$ is the fine structure constant. According to
the Weizs$\ddot{a}$cker-Williams\footnote{The
Weizs$\ddot{a}$cker-Williams (WW) approximation is a useful form
of approximation for the exact total cross-section for large
values of $\sqrt{s}$. In this approximation, the production rate
is described by a convolution of the probability for emitting a
photon from the electron with the corresponding real photon cross
section for the reaction $\gamma{g}{\rightarrow}t\overline{t}$
\cite{Baur}.} approximation \cite{WW1, WW2,WW3}, the total
cross-section for large values of $\sqrt{s}$ is provided by the
equation:
\begin{eqnarray} \label{Sigmaww eq}
\sigma=\int_{y_{\mathrm{min}}}^{1}dy
P_{\gamma}(y)\int_{z_{\mathrm{min}}}^{1}dz
xg(z,\mu_{F})\widehat{\sigma}.
\end{eqnarray}
Here, $P_{\gamma}$ is the photon-splitting function and
$\widehat{\sigma}$ is the cross section of the $t\overline{t}$
pair production by a real photon, where
$z_{\mathrm{min}}=\mu^{2}/(ys)$. The corresponding hadron-level
cross sections, $\sigma^{t}_{k=2,L}$ in accordance with the WW
approximation in the fixed-flavor-number scheme (FFNS), are
defined by the following form \cite{Schuler, Baur, Ivanov}
\begin{eqnarray} \label{Sigmakt eq}
\sigma^{t}_{k}=\int_{z_{\mathrm{min}}}^{1}dz
G(z,\mu_{F})\widehat{\sigma}^{t}_{k,g}(\frac{x}{z},\mu_{F},\mu_{R}),
\end{eqnarray}
where $\widehat{\sigma}^{t}_{k,g},~(k=2,L)$ are the $\gamma^{*}g$
cross sections of the photon -gluon component of the heavy-quark
leptoproduction at the LO approximation, given by:
\begin{eqnarray} \label{gtt eq}
\gamma^{*}+g{\rightarrow}t+\overline{t}.\nonumber
\end{eqnarray}
Here, $\mu_{R}$ and $\mu_{F}$ are the renormalization and
factorization scales, respectively. The default renormalisation
and factorization scales are set to be equal
$\mu_{R}^{2}=Q^{2}+4m_{t}^2$ and $\mu_{F}^{2}=Q^{2}$. The
leptoproduction cross sections $\sigma^{t}_{k=2,L}(x,Q^2)$ are
related to the DIS
structure functions $F^{t}_{k}(x,Q^{2})$.\\
The number of active flavors changes as the scale is increased,
therefore parton distributions are released in a variable-flavor-
number scheme (VFNS). Usually, $n_{f}=5$ is taken as the maximum
number of flavors by default. However, in NNPDF \cite{NNPDF},
$n_{f}=6$ PDF sets are also made available. This is important in
the running of $\alpha_{s}$ for precision phenomenology, as the
values of $\alpha_{s}$ with five and six active flavors already
differ by about $2\%$ at the TeV scale \cite{Rojo}. Investigations
into the top quark structure  inside the proton are also discussed
in \cite{LHeC, Boroun3, LHeC2, Kit}.\\
The total longitudinal and transverse cross sections in the
processes of top quark production in DIS via the one-photon
exchange mechanism are related to the corresponding DIS structure
functions as
\begin{eqnarray} \label{SigmaLT eq}
\sigma^{t}_{L}(x,Q^2)&=&\frac{4\pi^2\alpha_{em}}{Q^2}F^{t}_{L}(x,Q^{2}),\nonumber\\
\sigma^{t}_{T}(x,Q^2)&=&\frac{8\pi^2\alpha_{em}}{Q^2}xF^{t}_{1}(x,Q^{2}).
\end{eqnarray}
Instead of measuring $F_{1}$, experimentalists measure the
structure function as \cite{Van}
\begin{eqnarray} \label{F1 eq}
F^{t}_{2}(x,Q^{2})=2xF^{t}_{1}(x,Q^{2}) + F^{t}_{L}(x,Q^{2}).
\end{eqnarray}
The reduced cross-section for pair production of top quarks in
ep-collisions is defined in terms of the top quark structure
functions as
\begin{eqnarray} \label{Sigmareduced eq}
\sigma_{r}^{t}(x,Q^{2})=F_{2}^{t}(x,Q^{2})
-f(y)F_{L}^{t}(x,Q^{2}),
\end{eqnarray}
where $f(y)=y^{2}/1+(1-y)^{2}$ and $y$ represents the
inelasticity. In the collinear generalized double asymptotic
scaling (DAS) \cite{DAS1, DAS2, DAS3} approach, the top quark
structure functions are driven at small $x$ by gluons as
\begin{eqnarray} \label{Coif eq}
F_{k}^{t}(x,Q^{2})=e_{t}^{2}\sum_{n=0}a^{n+1}_{s}(Q^{2})B_{k,g}^{(n)}(x,a)
{\otimes} xg(x,Q^{2}),~~~(k=2, L),
\end{eqnarray}
where $a=m_{t}^{2}/Q^{2}$ and
$a_{s}(Q^{2})=\alpha_{s}(Q^{2})/4\pi$. The $\otimes$ symbol
denotes the convolution integral, which simplifies to a
multiplication in Mellin $N$-space. The notation is given by
$a(x)\otimes b(x)=\int_{x}^{1}\frac{dz}{z}a(z)b(\frac{x}{z})$ and
$xg(x,Q^{2})$ represents the gluon distribution function. The
coefficient functions at the Born level (first order in
$\alpha_{s}$) and the order $\alpha^{n}_{s}$ contributions are
reported in Refs.\cite{Kotikov1, Boroun4, Boroun5}. In order to address the
unphysical mass scale $\mu$, the renormalisation and factorisation
scale for the heavy quarks is defined as
$\mu^{2}=Q^{2}$ and $\mu^{2}=Q^{2}+4m_{t}^{2}$.\\
In this work, the collinear approach calculation of top quark pair
production at the  FCC-he and the LHeC kinematic regions is
presented. The focus is on the limited behavior of the reduced
cross section of top quark-pair production in the high
inelasticity region. The parametrization of gluon distribution and
dipole cross-section at the limit $x_{\mathrm{min}}=Q^2/s$ is
included. The probability of the production of the Higgs boson
from the pair top production in the $\gamma^{*}g$ domain of the ep
new colliders is defined. The rest of this paper is organized as
follows. In Sec. II, the theoretical framework for the coefficient
function in the collinear approach  is presented. In Sec. III, the
numerical results for the reduced cross section
$\sigma^{t\overline{t}}_{r}(Q^2/s,Q^{2})$ and for the dipole cross
section
 $\sigma_{\mathrm{dip}}^{t\overline{t}}(Q^2/s,Q^{2})$ at ultra-high energy processes are studied.\\

\subsection{II. Theory}
The top reduced cross-section, as given by Eq.~ (\ref{Sigmareduced eq}),
 can be rewritten in the following form in the limit $y=1$
\begin{eqnarray} \label{Sigmareducedy1 eq}
\mathrm{lim}_{y{\rightarrow}1}\Large{[}\frac{\sigma_{r}^{t}(Q^2/s,Q^{2})}{F_{2}^{t}(Q^2/s,Q^{2})}\Large{]}=1
-\frac{F_{L}^{t}(Q^2/s,Q^{2})}{F_{2}^{t}(Q^2/s,Q^{2})},
\end{eqnarray}
where $f(y)=1$ is defined at a fixed $\sqrt{s}$ and $Q^2$, and
$x_{\mathrm{min}}=Q^2/s$  dominates in nucleons and nuclei. This
limit indicates that the longitudinal polarization of the virtual
photon at $y=1$ approaches zero \cite{Taylor, Boroun6}. The ratio
of the DIS top structure functions corresponding to
$x_{\mathrm{min}}=Q^2/s$
 is defined by the following formula:
\begin{eqnarray} \label{FL2 eq}
F_{L2}^{t}(Q^2/s,Q^{2})=\frac{F_{L}^{t}(Q^2/s,Q^{2})}{F_{2}^{t}(Q^2/s,Q^{2})}=\frac{\sum_{n=0}a^{n+1}_{s}(\mu^{2})B_{L,g}^{(n)}(Q^2/s,a)
{\otimes}
xg(Q^2/s,Q^{2})}{\sum_{n=0}a^{n+1}_{s}(\mu^{2})B_{2,g}^{(n)}(Q^2/s,a)
{\otimes} xg(Q^2/s,Q^{2})}.
\end{eqnarray}
In  the collinear generalized DAS approaches, the coefficient
functions $B_{k,g}^{(n)}$ are known but the gluon distribution
functions have to be obtained from the Parton Distribution
Function groups. One can obtain a non-trivial upper bound on the
ratio $F_{L2}^{t}$  at the limit  $y=1$, which will be valid with
new data at the LHeC \cite{LHeC, LHeC2} and EIC \cite{EIC}.\\
We first consider the bound for the ratio $F_{L2}^{t}$ as we
define \cite{Ewerz, Prasz}
\begin{eqnarray} \label{Ratio eq}
g(Q^2,s,
m_{t})=\frac{\sum_{n=0}a^{n+1}_{s}(\mu^{2})B_{L,g}^{(n)}(Q^2/s,a)
}{\sum_{n=0}a^{n+1}_{s}(\mu^{2})B_{2,g}^{(n)}(Q^2/s,a)}=\frac{a^{1}_{s}(\mu^{2})B_{L,g}^{(0)}(Q^2/s,a)+a^{2}_{s}(\mu^{2})B_{L,g}^{(1)}(Q^2/s,a)+...
}{a^{1}_{s}(\mu^{2})B_{2,g}^{(0)}(Q^2/s,a)+a^{2}_{s}(\mu^{2})B_{2,g}^{(1)}(Q^2/s,a)+...}.
\end{eqnarray}
In Fig.1, the ratio $g(Q^2,s, m_{t})$ is shown at the LO and NLO
approximations due to the coefficients $n=0$ and $n=1$ in Eq.~
(\ref{Ratio eq}), respectively. In this figure we observe that the
results at the LO approximation have a bound limit for the LHeC
and FCC center-of-mass (COM) energies. The results for the LHeC
and FCC COM energies are very similar at the NLO approximation and
increase as $Q^2$ values increase. Indeed, a bound limit for the
DIS top structure functions is not visible for the LHeC and FCC
COM energies using the collinear approximation. Therefore, in the
following, we use the LO approximation using collinear
factorization.\\
According to Fig.1, we observe that the function $g(Q^2,s, m_{t}
)$ has a maximum at the LO approximation with ${\approx}0.288$ and
${\approx}0.259$ for the COM energies in the LHeC and FCC-eh
colliders. These values are comparable to the bounds in the color
dipole model (CDM) \cite{Ewerz, Prasz}.
\begin{figure}
\centering
\includegraphics[width=0.65\textwidth]{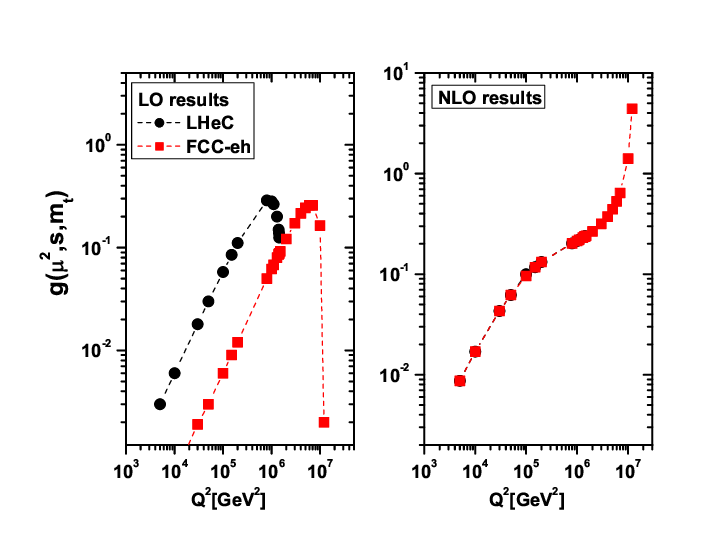}
\caption{The function $g(\mu^2,s, m_{t})$ is compared versus
$\mu^2=Q^2$ at the LO (left panel) and NLO (right panel)
approximations for the LHeC (black circles) and FCC-eh (red
squares) colliders with $\sqrt{s}=1.3$ and $3.5~\mathrm{TeV}$
respectively.}\label{Fig1}
\end{figure}
Because, the gluon density has to be non-negative, the bound
values lead to the bound of top structure functions as
\begin{eqnarray} \label{Bound eq}
\frac{F_{L}}{F_{2}}(Q^2,s, m_{t}){\leq} 0.259-0.288.
\end{eqnarray}
With respect to the bound of the ratio $F_{L2}^{t}$ (i.e.,
Eq.~(\ref{Bound eq})) at high inelasticity (i.e., $y=1$), the
bounds defined in the LHeC and FCC-eh COM energies do not cover
the gluon distributions at $x_{\mathrm{min}}$. Therefore, in the
following, we estimate the top reduced cross-section based on the
values
that support gluon density at low $x$ values.\\
The integrated gluon distribution function, $xg(x,Q^2)$, is
obtained from the unintegrated gluon distribution (UGD) in the
proton and nucleus using the correspondence between the color
dipole picture and the $k_{t}$-factorization formalism in the
leading logarithmic approximation \cite{Machado}. This method
covers kinematic ranges relevant for future lepton-hadron
colliders like LHeC/FCC-eh \cite{LHeC, FCC, LHeC2} and eRHIC
\cite{eRHIC}. In Ref.\cite{Golec1}, an analytical expression for
the gluon distribution function up to the saturation scale is
defined. The saturation scale marks the transition between the
dilute and saturated gluon system. It is defined as
$Q^{2}_{s}(x)=Q^{2}_{0}(x/x_{0})^{-\lambda}$ and is valid at
moderate $Q^2$ values. It is interesting to compare the upper
bound for the saturation scale from Fit.2 in Table 1 of
Ref.\cite{Golec2}. This plot shows the variation of $Q^{2}_{s}(x)$
for the LHeC and FCC-eh COM energies across different $Q^2$
values. In Fig.2, the behavior of the saturation scale at the
limit $x_{\mathrm{min}}$ is plotted as
$Q^{2}_{s}(x_{\mathrm{min}})=(\frac{\mu^2}{sx_{0}})^{-\lambda}~\mathrm{GeV}^2$
at the scales $\mu^2=Q^2$ and $\mu^2=Q^2+4m_{t}^{2}$. This plot
shows the variation of $Q^{2}_{s}(x_{\mathrm{min}})$ for the LHeC
and FCC-eh COM energies. As expected one finds that
$Q_{s}^{2}{\sim}1.61-2.63~\mathrm{GeV}^2$ in the low-$x$ region
where $x{\simeq}0.59{\times}10^{-6}$ and
$x{\simeq}0.82{\times}10^{-7}$ respectively. Moreover, we have
tested the sensitivity of the saturation scale with the top-quark
mass at the renormalization scale $\mu^2=Q^2+4m_{t}^{2}$, finding
a flat behavior for the saturation scale in the region
$10{\leq}Q^2<10^{5}~\mathrm{GeV}^2$. These findings at $\mu^2=Q^2$
confirm results at the high inelasticity within the color dipole
framework \cite{Machado2}.\\
\begin{figure}
\centering
\includegraphics[width=0.55\textwidth]{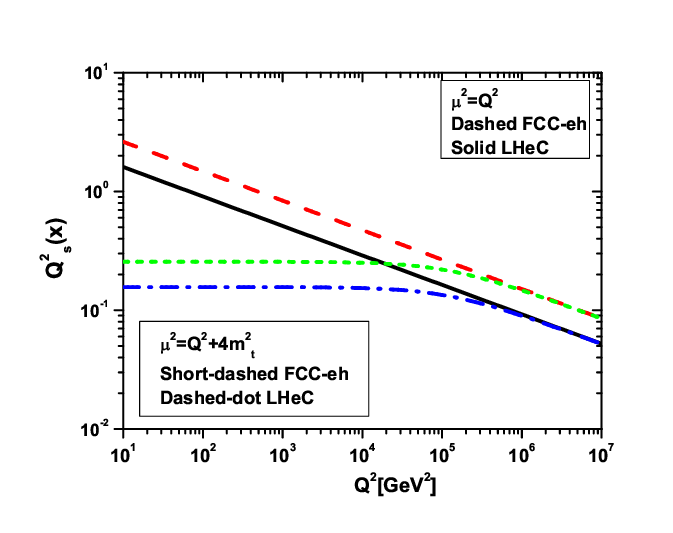}
\caption{Saturation scale at $x_{\mathrm{min}}$ is shown at
$\mu^2=Q^2$ and $\mu^2=Q^2+4m_{t}^{2}$ for the LHeC and FCC-eh COM
energies. LHeC: $\mu^2=Q^2$ (solid black curve) and
$\mu^2=Q^2+4m_{t}^{2}$ (dashed-dot blue curve); FCC-eh:
$\mu^2=Q^2$ (dashed red curve) and $\mu^2=Q^2+4m_{t}^{2}$
(short-dashed green curve).}\label{Fig1}
\end{figure}
The UGD in the small-$x$ regime is determined by the following
formula\footnote{The relationship between the integrated and
unintegrated gluon distribution functions at very low $x$ is
defined as follows: $$
f_{g}(x,k_{t}^2){\simeq}\frac{\partial}{\partial{\mu^2}}G(x,\mu^2)\bigg|_{\mu^2=k_{t}^2},
$$\\
which can be adjusted by incorporating the Sudakov form factor
\cite{SUD1, SUD2,SUD3}
$$
f_{g}(x,k_{t}^2,\mu^2){\simeq}\frac{\partial}{\partial{k_{t}^2}}[T_{g}(k_{t}^2,\mu^2)G(x,k_{t}^2)].
$$\\
In this context, the resummation of large Sudakov logarithms like
$\log(k_{t}/Q)$, as well as $\log(1/x)$, is applicable in the
region where $Q_{s}{\sim}k_{t}{\ll}P_{t}{\sim}Q$. }
\begin{eqnarray} \label{UGD eq}
\alpha_{s}\mathcal{F}(x,k_{t})=\frac{3\sigma_{0}}{4\pi^2}(k^{2}_{t}/Q^{2}_{s})
\exp(-k^{2}_{t}/Q^{2}_{s}).
\end{eqnarray}
The dipole cross section is related to the UGD through a Fourier
transform \cite{Nikolaev1, Nikolaev2}. Various models
\cite{Nikolaev3, Ivanov2, Vera, Watt, Golec3} have been developed
to establish the relationship between the UGD and the $k_{t}$
factorization. The $k_{t}$ factorization approach to the
production of $\phi$ and $\rho$ mesons in DIS via the
helicity-conserving $\gamma^{*}(T,L){\rightarrow}V$ impact factor
is discussed in Ref.\cite{Celebrato1, Celebrato2, Celebrato3}. The
results of cross sections for longitudinally and transversely
polarized mesons presented at $x_{\mathrm{min}}=Q^2/s$ strongly
depend on the choice of UGD. In Fig.3, we plot the  $k_{t}^{2}$
dependence of the UGD at $x_{\mathrm{min}}=Q^2/s$ with respect to
the LHeC  and FCC-eh COM energies. We observe that there is a
maximum value for $f(x_{\mathrm{min}},k_{t}^{2})$ at $\mu^2=Q^2$
and $\mu^2=Q^2+4m_{t}^{2}$. These maximum values increase with
increasing $Q^2$ values when $\mu^2=Q^2$ is changed to
$\mu^2=Q^2+4m_{t}^{2}$, and decrease to low $k_{t}^{2}$ as $Q^2$
increases. An enhancement followed by a depletion in the UGD
behavior at $x_{\mathrm{min}}=Q^2/s$ is observable in the LHeC and
FCC-eh COM energies with increasing values of $k_{t}$
\cite{Boroun7, Boroun8}.\\
\begin{figure}
\centering
\includegraphics[width=0.65\textwidth]{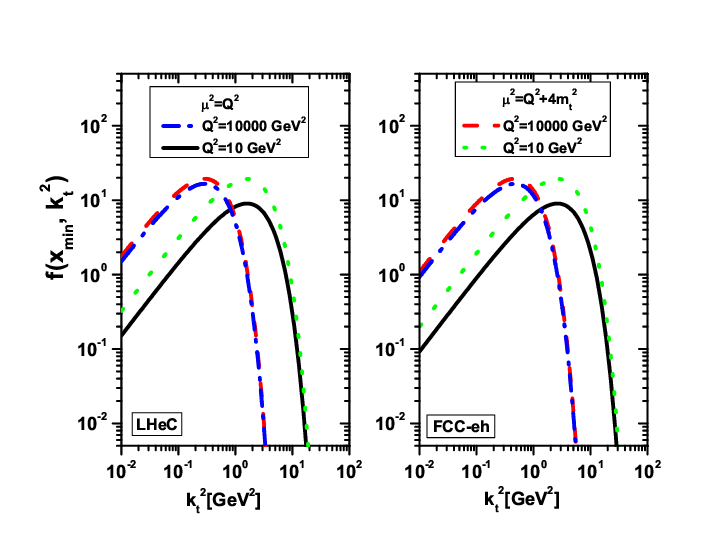}
\caption{UGD obtained at $x_{\mathrm{min}}$ is shown at
$\mu^2=Q^2$ ($Q^2=10~\mathrm{GeV}^2$ (solid black curve) and
$Q^2=10000~\mathrm{GeV}^2$ (dashed-dot blue curve)) and
$\mu^2=Q^2+4m_{t}^{2}$ ($Q^2=10~\mathrm{GeV}^2$ (dot green curve)
and $Q^2=10000~\mathrm{GeV}^2$ (dashed red curve)) for the LHeC
(left panel) and FCC-eh (right panel) COM energies as a function
of $k^2_{t}$.}\label{Fig3}
\end{figure}
The integrated gluon distribution, as defined by the UGD equation (i.e.,
Eq.~(\ref{UGD eq})), can be expressed in the following form
\cite{Golec1, Boroun7}
\begin{eqnarray} \label{Gluon eq}
xg(x,Q^2)=\int_{0}^{Q^2}dk^{2}_{t}\mathcal{F}(x,k_{t})=
\frac{3\sigma_{0}}{4\pi^2\alpha_{s}}Q^{2}_{s}\bigg{[}
1-\bigg{(}1+\frac{Q^2}{Q^{2}_{s}}\bigg{)}e^{-\frac{Q^2}{Q^{2}_{s}}}\bigg{]},
\end{eqnarray}
where the coefficients of the model are defined in
Ref.\cite{Golec2}.
 The use of the gluon distribution function in heavy quark
structure functions in quantum chromodynamics (QCD) is dominant in
the fusion reaction $t\overline{t}$ production at high-energy
colliders, which is the direct detection of entanglement
\cite{Afik}.\\
The dipole cross section of the Bartels-Golec-Biernat-Kowalski
(BGK) model\footnote{The dipole cross section in the BGK model
is directly related to the gluon distribution and indirectly 
dependent on the UGD. The UGD, through a Fourier transform, is
related to the gluon distribution according to the Eqs.~(\ref{Dipoler
eq}), (\ref{Dipole eq}), (\ref{Gluon eq}), and (\ref{UGD eq}).}
\cite{BGK} at $x_{\mathrm{min}}=\mu_{r}^2/s$ ia as follows:
\begin{eqnarray} \label{Dipole eq}
\sigma_{\mathrm{dip}}(\mu_{r}^2/s,r)=\sigma_{0}\bigg{\{}1-\exp{\bigg{(}}\frac{-\pi^2r^2\alpha_{s}(\mu_{r}^2)xg(\mu_{r}^2/s,\mu_{r}^2)}{3\sigma_{0}}\bigg{)}\bigg{\}},
\end{eqnarray}
where the evolution scale $\mu_{r}^{2}$ is connected to the size
$r$ of the dipole by $\mu_{r}^{2}=C/r^2+\mu_{0}^{2}$ and the
coefficients due to the upper bound are selected from Fit.2 in
Table 3 of Ref.\cite{Golec2}. In the color dipole picture (CDP),
the virtual photon $\gamma^{*}$ dissociates into a $t\overline{t}$
dipole-pair which interacts with the color fields in the proton.
The imaginary part of the $(t\overline{t})p$ forward scattering
amplitude is defined by the dipole cross-section,
$\sigma_{\mathrm{dip}}(\mu_{r}^2/s,r)$ in the following form
\begin{eqnarray} \label{Sigmadip eq}
\sigma_{L,T}^{\gamma^{*}p}(\mu_{r}^2/s,\mu_{r}^2)=\frac{1}{4\pi}{\int}dzd^2\mathbf{r}|\Psi_{L,T}(r,z,\mu_{r}^2)|^{2}
\sigma_{\mathrm{dip}}(\mu_{r}^2/s,r),
\end{eqnarray}
where the $\gamma^{*}p$ cross-sections are related to the DIS
structure functions. Indeed, the DIS total scattering cross
sections $\sigma_{L,T}(x,\mu^2)$ can be defined in QCD as an
effective theory in the high-energy limit. This theory is known as
the color glass condensate effective field theory, based on the
knowledge of the incoming and outgoing elementary particles
\cite{CGC1, CGC2, CGC3, CGC4}.  The longitudinal momentum
fractions carried by the dipole are $z$ and $1-z$, and the virtual
wave function of the photon is defined by $\Psi_{L,T}$ according
to the longitudinal and transverse photon polarizations. The
kernel $|\Psi|^2$ of the light-cone wave functions in
Eq.~(\ref{Sigmadip eq}) based on the light-cone perturbative
theory is
\begin{eqnarray} \label{PsiT eq}
|\Psi^{\gamma^{*}{\rightarrow}t\overline{t}}_{T}(r,z,\mu_{r}^2)|^{2}=\frac{2N_{c}\alpha_{\mathrm{em}}e_{t}^{2}}{\pi}
(m_{t}^{2}K_{0}^{2}(\epsilon_{t}r)+[z^2+(1-z)^2]\epsilon_{t}^{2}K_{1}^{2}(\epsilon_{t}r)),
\end{eqnarray}
and
\begin{eqnarray} \label{PsiL eq}
|\Psi^{\gamma^{*}{\rightarrow}t\overline{t}}_{L}(r,z,\mu_{r}^2)|^{2}=\frac{8N_{c}\alpha_{\mathrm{em}}e_{t}^{2}}{\pi}
\mu^2z^2(1-z)^2K_{0}^{2}(\epsilon_{t}r),
\end{eqnarray}
where $N_{c}$ is the number of color charges in QCD and $K_{i}$
are Bessel functions of the second kind. Here,
$\epsilon_{t}=z(1-z)\mu^2+m_{t}^2$. The effective dipole
cross-section at high inelasticity is evaluated in the following
form
\begin{eqnarray} \label{Dipoler eq}
\sigma_{\mathrm{dip}}(\mu_{r}^2/s,r)/\sigma_{0}=1-\exp{\bigg{(}}\frac{-r^2}{4}\mu^{2}_{s}\bigg{[}
1-\bigg{(}1+\frac{\mu^2}{\mu^{2}_{s}}\bigg{)}e^{-\frac{\mu^2}{\mu^{2}_{s}}}\bigg{]}\bigg{)},
\end{eqnarray}
where
$\mu^{2}_{s}=(\frac{\mu^2}{sx_{0}})^{-\lambda}~\mathrm{GeV}^2$,
which depends on the the COM energies and the top-quark effective
mass when $\mu^2=\mu_{r}^2$ and $\mu^2=\mu_{r}^2+4m_{t}^2$. In
Fig.4, the results of the dipole cross-sections,
$\sigma_{\mathrm{dip}}(\mu_{r}^2/s,r)/\sigma_{0}$, of the BGK
model at $x_{\mathrm{min}}$ are shown as a function of $r$ at
scales $\mu^2=\mu_{r}^2$ and $\mu^2=\mu_{r}^2+4m_{t}^{2}$ for the
LHeC and FCC-eh COM energies. This behavior for the top quark mass
is bounded based on the coefficients in the CDP. We observe that
$\sigma_{\mathrm{dip}}/\sigma_{0}$ is plotted as a function of the
dipole size $r$. At the scale $\mu^2=\mu_{r}^2+4m_{t}^{2}$ the
behavior at the LHeC and FCC-eh COM energies is similar to the GBW
model \cite{Golec1, Golec2, Golec3}. The dipole cross-sections at
the scale $\mu^2=\mu_{r}^2$ show  a violation for
$3{\times}10^{-2}{\lesssim}r{<}1~\mathrm{fm}$, and they exhibit
the same behavior at the scale $\mu^2=\mu_{r}^2+4m_{t}^{2}$ for
the regions $r{\gtrsim}1~\mathrm{fm}$ and
$r{\lesssim}3{\times}10^{-2}~\mathrm{fm}$.\\
\begin{figure}
\centering
\includegraphics[width=0.55\textwidth]{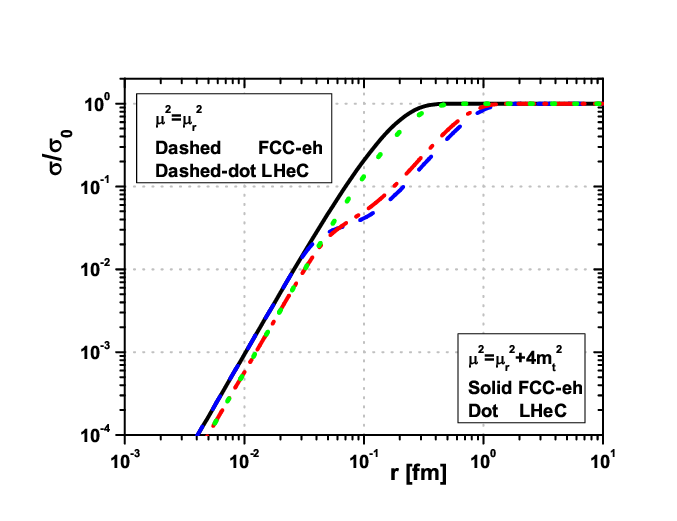}
\caption{The extracted ratio
$\sigma_{\mathrm{dip}}(\mu_{r}^2/s,r)/\sigma_{0}$ as a function of
$r$ at $x_{\mathrm{min}}$ is shown at $\mu^2=\mu_{r}^2$ and
$\mu^2=\mu_{r}^2+4m_{t}^{2}$ for the LHeC and FCC-eh COM energies.
LHeC: $\mu^2=\mu_{r}^2$ (dashed-dot red curve) and
$\mu^2=\mu_{r}^2+4m_{t}^{2}$ (dot green curve); FCC-eh:
$\mu^2=\mu_{r}^2$ (dashed blue curve) and
$\mu^2=\mu_{r}^2+4m_{t}^{2}$ (solid black curve).}\label{Fig4}
\end{figure}
The DIS structure functions can be defined as an integral
transform of the dipole cross section against the kernel
$r\mathcal{Z}(r,\mu^2)$, which is defined by the following form
\cite{Hann}
\begin{eqnarray} \label{FTL eq}
F_{T,L}(\mu^2/s,\mu^2)=\int_{0}^{\infty}\mathcal{Z}_{T,L}(r,\mu^2)\sigma_{\mathrm{dip}}(\mu_{r}^2/s,r)rdr,
\end{eqnarray}
where the $z$-integrated wave function kernel is
\begin{eqnarray} \label{Zkernel eq}
\mathcal{Z}_{T,L}(r,\mu^2)=\frac{\mu^2}{4\pi\alpha_{\mathrm{em}}}{\int}_{0}^{1}|\Psi^{\gamma^{*}{\rightarrow}t\overline{t}}_{T,L}(r,z,\mu_{r}^2)|^{2}dz.
\end{eqnarray}
The ratio $\sigma_{r}/F_{2}$ at high inelasticity is defined by
the following form
\begin{eqnarray} \label{SigmaF2CGC eq}
\frac{\sigma_{r}}{F_{2}}(\mu^2/s,\mu^2)=\frac{F_{T}}{F_{T}+F_{L}}(\mu^2/s,\mu^2)
{\simeq}1-\frac{F_{L}}{F_{T}}(\mu^2/s,\mu^2)+(\frac{F_{L}}{F_{T}}(\mu^2/s,\mu^2))^2+...,
\end{eqnarray}
where
\begin{eqnarray} \label{FTLZ eq}
\frac{F_{L}}{F_{T}}(\mu^2/s,\mu^2)=\frac{\int_{0}^{\infty}\widetilde{\mathcal{Z}}_{L}(r,\mu^2)\sigma_{\mathrm{dip}}(\mu_{r}^2/s,r)dr}
{\int_{0}^{\infty}\widetilde{\mathcal{Z}}_{T}(r,\mu^2)\sigma_{\mathrm{dip}}(\mu_{r}^2/s,r)dr},
\end{eqnarray}
with $\widetilde{\mathcal{Z}}{\equiv}r{\mathcal{Z}}$, where the
DIS structure functions in Eq.~(\ref{FTLZ eq}) are written into a
linear integral transformation of the dipole cross section. The
ratio $\frac{\sigma_{r}}{F_{2}}(\mu^2/s,\mu^2)$ at the limit
$x_{\mathrm{min}}$ in the dipole size range
$0.0001<r<100~\mathrm{fm}$ has a bound in the range of
${\simeq}0.980$ and $0.978$ for $\mu^2=\mu_{r}^2$ and
$\mu^2=\mu_{r}^2+4m_{t}^{2}$ respectively at the LHeC and FCC-eh
COM energies.\\
The LHeC and FCC-eh will explore the strange density and momentum
fraction carried by top quarks, as the total cross-section of top-
antitop quark pairs $(t\overline{t})$ at photoproduction is
expected to be $0.05~\mathrm{pb}$ at the LHeC \cite{LHeC}. The CP
nature of the top-antitop-Higgs coupling can be analysed at the
LHeC in $ep{\rightarrow}et\overline{t}H$ production \cite{LHeC,
Kumar}. The observation of $t\overline{t}H$ production is based on
a combined analysis of proton-proton collision data at COM
energies of $\sqrt{s}=7, 8$, and $13~\mathrm{TeV}$ for a Higgs
boson mass of $125.09~\mathrm{GeV}$, as reported in
Ref.\cite{Higgs5}. Recently, the most model-independent direct
measurement was reported in Ref.\cite{ATLAS1}, with sensitivity to
the top-quark Yukawa coupling. Higgs production via vector-boson
fusion at the
$\gamma^{*}g{\rightarrow}t\overline{t}(t\overline{t}{\rightarrow}H)$
will be discussed in the LHeC and FCC-eh colliders. The
cross-section for this process is similar to the
$gg{\rightarrow}t\overline{t}(t\overline{t}{\rightarrow}H)$ as
reported in Refs.\cite{Fabio1, Fabio2} where the top quark loop is
connected with the Higgs by the coupling
$m_{t}(1+\frac{H}{\upsilon})t\overline{t}$, where
$\upsilon{\approx}246~\mathrm{GeV}$ is the vacuum expectation
value of the Higgs field ($H$). These fields naturally have a mass
above $2m_{t}$ as the top quark has a large Yukawa coupling with
the Higgs boson \cite{Roberto}. Indeed, measuring Higgs boson
production in association with a $t\overline{t}$ pair is crucial
for understanding the top-quark couplings to the Higgs boson at
the LHeC and FCC-eh COM energies compared to
beyond-the-Standard-Model(BSM) theories \cite{Ana, Khanpour}. The
cross-section for Higgs production from gluon fusion at the LHeC
and FCC-eh colliders is approximately defined as:
\begin{eqnarray} \label{Higgs eq}
\sigma(\gamma^{*}g{\rightarrow}t\overline{t}(t\overline{t}{\rightarrow}H))<\frac{\alpha_{em}}{\alpha_{s}}\sigma(gg{\rightarrow}t\overline{t}(t\overline{t}{\rightarrow}H)).
\end{eqnarray}
\section{III. Results and Discussion}

Determination of the strong coupling constant $\alpha_{s}(m_{z})$
from deep-inelastic scattering and hadron collider data without a
simultaneous determination of the PDFs leads to an explicit test
of the partial $\alpha_{s}$ value obtained from the top quark pair
production data. This value is reweighted by a factor of
$N^{\mathrm{tot}}_{\mathrm{dat}}/N^{t\overline{t}}_{\mathrm{dat}}$
and is estimated to be:
\begin{eqnarray} \label{Alphat eq}
\alpha_{s}^{t\overline{t}}(m_{z})=0.1201{\pm}0.0012,
\end{eqnarray}
as reported in Ref.\cite{Forte}. In the following (i.e.,
Figs.5-8), the DIS top structure functions at the limit
$x_{\mathrm{min}}=Q^2/s$ in Eq.~(\ref{FL2 eq})  depend on
the gluon distribution function of Eq.~(\ref{Gluon eq}) which is
obtained from the UGD.\\
 In Fig.5, we show the behavior of
the ratio of top DIS structure functions, Eq.~(\ref{FL2 eq}), at
the limit of $x_{\mathrm{min}}=Q^2/s$ for $\sqrt{s}=1.3$ and
$3.5~\mathrm{TeV}$ at the scales $\mu^2=Q^2$ and
$\mu^2=Q^2+4m_{t}^{2}$ respectively. In Fig.5, we observe that the
ratio given by Eq.~(\ref{FL2 eq}) for $F_{L2}$  is equal to the
bound specified by Eq.~(\ref{Bound eq}) within the range of
$10^{5}<Q^2<10^{7}~\mathrm{GeV}^2$. This demonstrates the
importance of the longitudinal structure function of top-pair
production in the LHeC and FCC-eh colliders at high inelasticity.
The ratio $F_{L2}$ shows a moderate slope at
$Q^2<10^{5}~\mathrm{GeV}^2$ in the scale $\mu^2=Q^2+4m_{t}^2$. The
data at small $Q^2$ values will span a small range of $x$, and
this range varies strongly with $\mu^2$. The importance of the
longitudinal structure function at large $Q^2$ values is shown in
Fig.6. In Fig.6, the ratio $\sigma_{r}/F_{2}$ at the limit of
$x_{\mathrm{min}}=Q^2/s$ for $\sqrt{s}=1.3$ and $3.5~\mathrm{TeV}$
at the scales $\mu^2=Q^2$ and $\mu^2=Q^2+4m_{t}^{2}$ respectively
is shown. We observe that the longitudinal structure function for
top-quark pair production at $Q^2<100~\mathrm{GeV}^2$ is equal to
zero. Therefore the bound in Eq.~(\ref{Sigmareduced  eq}) for
$\sigma_{r}/F_{2}$ is equal to 1 at low $Q^2$ values for both
scales $\mu^2=Q^2$ and $\mu^2=Q^2+4m_{t}^{2}$, as we observe
\begin{eqnarray} \label{Sigmaratio eq}
\mathrm{lim}_{y{\rightarrow}1}\Large{[}\frac{\sigma_{r}^{t}(\mu^2/s,\mu^{2})}{F_{2}^{t}(\mu^2/s,\mu^{2})}\Large{]}{\simeq}1.
\end{eqnarray}
This means that at low $Q^2$ values (i.e.,
$Q^2<100~\mathrm{GeV}^2$)
$\sigma_{r}^{t}(\mu^2/s,\mu^{2}){\simeq}F_{2}^{t}(\mu^2/s,\mu^{2})$,
which determines the transversely
polarized scattering at low $Q^2$ values.\\
\begin{figure}
\centering
\includegraphics[width=0.55\textwidth]{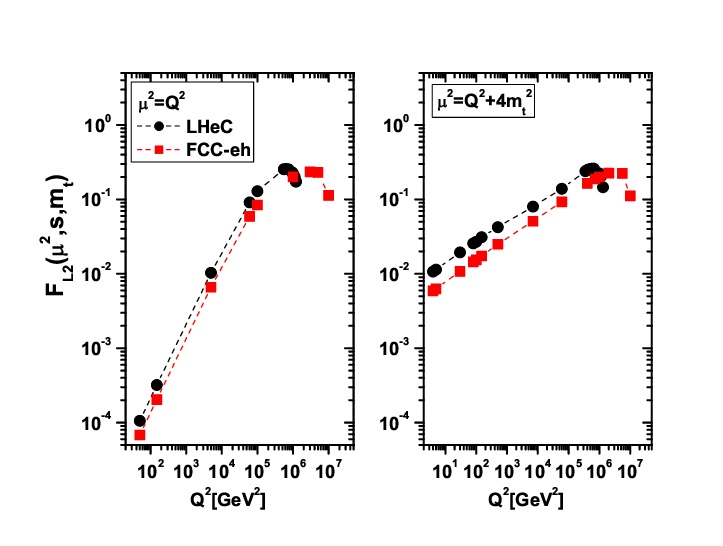}
\caption{The ratio
$F_{L2}^{t}{\equiv}\frac{F_{L}^{t}}{F_{2}^{t}}(\mu^2,s,m_{t})$ as
a function of $Q^2$ at $x_{\mathrm{min}}$ is shown at $\mu^2=Q^2$
(left panel) and $\mu^2=Q^2+4m_{t}^{2}$ (right panel) for the LHeC
(black circles) and FCC-eh (red squares) COM energies with
$\sqrt{s}=1.3$ and $3.5~\mathrm{TeV}$ respectively.}\label{Fig5}
\end{figure}
\begin{figure}
\centering
\includegraphics[width=0.55\textwidth]{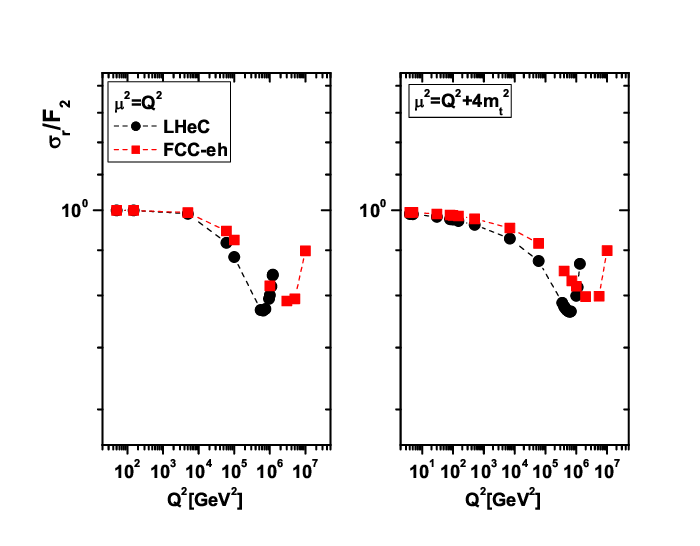}
\caption{ The ratio
$\frac{\sigma_{r}^{t}}{F_{2}^{t}}(\mu^2,s,m_{t})$ as a function of
$Q^2$ at $x_{\mathrm{min}}$ is shown at $\mu^2=Q^2$ (left panel)
and $\mu^2=Q^2+4m_{t}^{2}$ (right panel) for the LHeC (black
circles) and FCC-eh (red squares) COM energies with $\sqrt{s}=1.3$
and $3.5~\mathrm{TeV}$ respectively. }\label{Fig6}
\end{figure}
\begin{figure}
\centering
\includegraphics[width=0.55\textwidth]{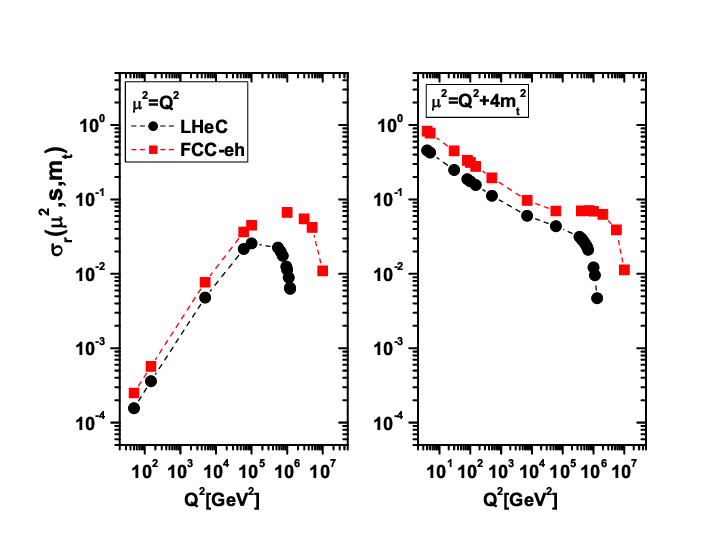}
\caption{ The top reduced cross-section as a function of $Q^2$ at
$x_{\mathrm{min}}$ is shown at $\mu^2=Q^2$ (left panel) and
$\mu^2=Q^2+4m_{t}^{2}$ (right panel) for the LHeC (black circles)
and FCC-eh (red squares) COM energies with $\sqrt{s}=1.3$ and
$3.5~\mathrm{TeV}$ respectively. }\label{Fig7}
\end{figure}
In Fig.7, we display the behavior of the top reduced
cross-section, $\sigma_{r}^{t}(\mu^2/s,\mu^{2})$ at the limit of
$x_{\mathrm{min}}=Q^2/s$ for $\sqrt{s}=1.3$ and $3.5~\mathrm{TeV}$
at the scales $\mu^2=Q^2$ and $\mu^2=Q^2+4m_{t}^{2}$. These
behaviors versus of $Q^2$ are observable in the COM energies of
new colliders. We observe that at $Q^2<100~\mathrm{GeV}^2$ (i.e.,
low $x$ values ), the top reduced cross sections (and top
structure functions) will be visible in the ranges
$10^{-4}<\sigma_{r}^{t}<10^{-3}$ at the scale $\mu^2=Q^2$ and also
$10^{-1}<\sigma_{r}^{t}<1$ at the scale $\mu^2=Q^2+4m_{t}^{2}$ for
the LHeC and FCC-eh  COM energies. We observe that the theoretical
uncertainty related to the freedom in the choice of $\mu$ (i.e.,
$\mu^2=Q^2$ and $\mu^2=Q^2+4m_{t}^{2}$) is negligibly small at
both the LHeC and FCC-eh COM energies at large $Q^2$ values (i.e.,
$Q^2{\gtrsim}4m_{t}^{2}$). However, at small $Q^2$ values, the
uncertainties between the results are large ,as shown in Fig.7.\\
The behavior of the top reduced cross-section at the scale
$\mu^2=Q^2+4m_{t}^{2}$ at both COM energies is modified by
applying the modification of the Bjorken scaling behavior. In a
case where the photon wave function depends on the mass of the
heavy quarks for low $Q^2$, one must take into account the
modification of the Bjorken variable $x$ in high inelasticity in
the following form:
\begin{eqnarray} \label{BJM eq}
x=\frac{Q^2}{s}+\frac{4m_{t}^{2}}{s}.
\end{eqnarray}
\begin{figure}
\centering
\includegraphics[width=0.55\textwidth]{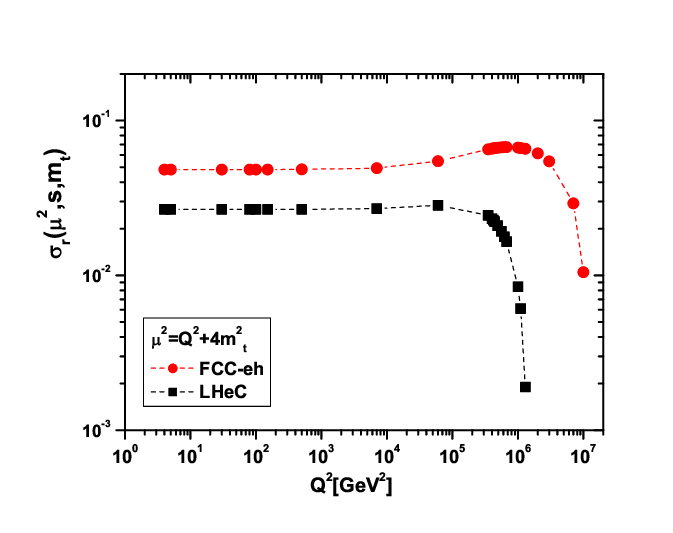}
\caption{ The top reduced cross-section as a function of $Q^2$ at
 $x=\frac{\mu^2}{s}$ is
shown at  $\mu^2=Q^2+4m_{t}^{2}$  for the LHeC (black circles) and
FCC-eh (red squares) COM energies. }\label{Fig8}
\end{figure}
In order to assess the significance of the mass $t\overline{t}$
pair quarks in the modified Bjorken variable, we show in Fig.8 the
mass dependence of the top reduced cross section at both the LHeC
and FCC-eh COM energies with $\mu^2=Q^2+4m_{t}^{2}$. In Fig.8, we
observe that the quality of the top reduced cross-section is
better if one includes the top quark mass in the definition of the
scaling variable, as the top reduced cross-section increases as
$Q^2$ increases. One can see that the top reduced cross-sections
obtained for high inelasticity with the modified Bjorken variable
are smaller and flatter than those without the effects of the top
quark mass. For $Q^2{\lesssim}4m_{t}^{2}$, we observe that
$\sigma_{r}^{t}$$^{,}s$ are approximately 0.027 and 0.049 with
$\sqrt{s}=1.3$ and $3.5~\mathrm{TeV}$ respectively. On the other
hand, for $Q^2{>}4m_{t}^{2}$, the behavior of $\sigma_{r}^{t}$
with and without the modification of the Bjorken variable is
similar. Therefore, it found to be approximately $Q^2$ independent
in the small-$Q^2$ regime (i.e., $Q^2{\lesssim}4m_{t}^{2}$) as the
saturation scale $Q_{s}^{2}$ (i.e., Fig.2) has this behavior at
the scale $\mu^2=Q^2+4m_{t}^{2}$. Then it continues to rise with
$Q^2$ until reach maxima, beyond which it fall for
$Q^2{\gg}4m_{t}^{2}$.\\

In conclusion, we have analyzed the behavior of the DIS structure
function of top quark pair production at $y=1$, where
$x_{\mathrm{min}}=Q^2/s$, in both LHeC and FCC-eh kinematics. The
ratio $\frac{F_{L}}{F_{2}}(\mu^2/s,\mu^2)$, considering the effect
of the gluon density due to the UGD at the renormalization scales,
is examined. This ratio with a limited bound derived in the
framework of the collinear approach, provides valuable information
on the dipole picture.
 The dipole cross section for the production of
$t\overline{t}$ in the BGK model depends on the renormalization
scales at a moderate dipole size $r$. We demonstrate that the
approximation $\sigma_{r}^{t}(\mu^2,s,m_{t}^{2}){\simeq}
F_{2}^{t}(\mu^2,s,m_{t}^{2})$ is dominant for $Q^{2}<4m_{t}^{2}$
and the saturation of the dipole cross-section is valid in this
domain. Finally, the top reduced cross-sections at
$x_{\mathrm{min}}$ for the LHeC and FCC-eh COM energies are
calculated and the effects of the modification of the Bjorken
scaling are applied at $\mu^2=Q^2+4m_{t}^{2}$. As predicted by the
authors in Ref.\cite{Baur}, a TeV ep-collider (i.e., LHeC and
FCC-eh colliders) will provide the exact pair production rate into
WW approximation. Therefore, the LHeC and FCC-eh should certainly
 consider the longitudinal structure function for top
production at $Q^2{\gtrsim}4m_{t}^{2}$ as an important item on the
list of physics topics.\\



\end{document}